# Building an Inclusive AAS - The Critical Role of Diversity and Inclusion Training for AAS Council and Astronomy Leadership

Brinkworth, C. S.[1]; Byrd Skaer, A. M. [2]; Prescod-Weinstein, C[3];, Teske, J.[4]; Tuttle S.[5]

[1]National Center for Atmospheric Research (NCAR), Boulder, CO; [2]Cornell University, Ithaca, NY; [3]Department of Physics, University of Washington, Seattle, WA; [4]Carnegie Institution for Science, Washington DC; [5]Department of Astronomy, University of Washington, Seattle, WA

Diversity, equity and inclusion are the science leadership issues of our time. As our nation and the field of astronomy grow more diverse, we find ourselves in a position of enormous potential and opportunity: a multitude of studies show how groups of diverse individuals with differing viewpoints outperform homogenous groups to find solutions that are more innovative, creative, and responsive to complex problems, and promote higher-order thinking amongst the group (Antonio et al, 2004; Page, 2007; Sommers, 2007). Research specifically into publications also shows that diverse author groups publish in higher quality journals and receive higher citation rates (Freeman & Huang, 2014). As we welcome more diverse individuals into astronomy, we therefore find ourselves in a position of potential never before seen in the history of science, with the best minds and most diverse perspectives our field has ever seen.

Despite this enormous growing potential, and the proven power of diversity, the demographics of our field are not keeping pace with the changing demographics of the nation. According to the United States Census, in July 2015 13.3% of Americans identified as Black or African American alone, 17.6% identified as Hispanic or Latino alone, and 1.2% identified as American Indian or Alaska Native alone; 2.6% of citizens identified as two or more races. The Census Bureau also predicts that by 2044, whites will no longer be a racial majority in the United States (Colby & Ortman, 2015). Right now, only 2.1% of astronomers identify as Black or African-American and 3.2% as Hispanic, Latina/o, or of Spanish origin and extremely few are Native or indigenous (Ivie et al., 2014). Disappointingly, these numbers for Physics and Astronomy have remained essentially constant between 2004 and 2012. This underrepresentation is most acute in leadership roles and on the key committees that shape the future of our field. When considering those with multiple minority identities, the level of underrepresentation is compounded: in 2012, there were fewer than 75 faculty members in Physics or Astronomy in the United States who are both female and African-American or Hispanic (Ivie et al., 2014).

Much effort has been put into increasing the pipeline of underrepresented students into the sciences, but there is significant evidence to suggest that many minoritized students enter the STEM fields and then drop out before graduation. We can compare the number of freshmen in 2008 who reported an intent to major in STEM to the actual number of STEM degrees awarded in 2012 (Table 1; National Science Foundation, 2015; Brinkworth, 2016) and we find that there are huge disparities between graduation rates of White and Asian males compared to minoritized groups. There are many reasons that students drop out of STEM, including mismatches between interest and choice of degree, as well as financial issues (Seymour & Hewitt, 1997), but there is a wealth of research to suggest that these disparities in graduation rates across race and gender are largely due to institutional barriers and "chilly" or "hostile" climates that underrepresented groups experience in STEM fields. Research also suggests that work to remove those barriers and create more welcoming climates could dramatically increase retention rates (Seymour  Hewitt, 1997; Yosso et al. 2009; Cech & Waidzunas, 2011; Fouad et al. 2012).

| Table 1: Percentage of freshman intending to major in STEM in 2008, who received STEM degrees in 2012. Source: National Science Foundation, 2015; Brinkworth, 2016 | | | |
|---|---|---|---|
| | **All** | **Male** | **Female** |
| **All** | 25.3 | 27.8 | 23.0 |
| **White** | 28.4 | 31.1 | 25.9 |
| **Asian/Pacific Islander** | 31.2 | 32.1 | 31.9 |
| **Black** | 14.0 | 16.3 | 12.8 |
| **Hispanic** | 16.9 | 17.1 | 16.3 |
| **American Indian/Alaska Native** | 16.1 | 17.9 | 15.6 |

There is a strong temptation to not discuss issues of race, gender identity, sexual orientation, and ableness in science. Many practitioners are uncomfortable with these topics, and believe that discussions around identity are irrelevant in a data-driven field like astronomy, where the scientific ideal is a meritocracy. However, astronomy has an existing culture, driven by "norms" that were established when our field was far less diverse. We continue to expect students, staff and faculty from all backgrounds to assimilate to the existing culture, rather than adapting that culture to become fully inclusive (Bell et al. 2009; Lee, 2015). This creates a heavy burden for those who are not part of that existing culture: Seymour & Hewitt (1997) found evidence that retention of minoritized groups in STEM is unrelated to their aptitude, but highly dependent on their ability to navigate the difficult social aspects of the fields. In a study of gender experiences in workplaces, Eisenhart & Finkel (1998) found that this cultural streamlining towards the dominant group can be invisible to all groups: both men and women in the study reported equal treatment, despite the researchers' objective observations of inequality in the workplace in favor of the male students and employees. Johnson (2007) found that a belief in "colourblind" and "gender-blind" meritocracy can negatively affect non-white, non-male students, reporting that:

> "*This match between Whiteness, maleness, and the characteristics needed for success in science was hidden in this setting by the silence about race, ethnicity, and gender, which was in turn hidden by the rhetoric of meritocracy. This silence prevented students and professors from seeing how ethnic, racial, and gendered dynamics helped determine which students found it easier to thrive.*"

Far from ensuring an equal, meritocratic workplace for all, a refusal to discuss social aspects of the STEM culture can be extremely stressful to underrepresented students, who are already being required to do extensive emotional work to navigate these unwelcoming climates (e.g. Seymour & Hewitt, 1997; Yosso et al, 2009; Fouad et al. 2012). In many cases, the silence leads to worse academic and mental and physical health outcomes for those we should be nurturing and celebrating as they start their science careers (Meyer, 1995; Meyer, 2003; Huebner & Davis, 2007; Nadal et al. 2011; Bockting et al. 2013).

Likewise, a refusal to discuss how discrimination in classrooms and the workplace impact specific groups of underrepresented minorities results in further amplifying the physical, emotional, academic and professional harm done to individuals within that specific group. Notably, the pervasive and uniquely targeted violence against African Americans has become highly visible over the last five years. For this reason, it is critical that the leaders of our field explicitly and openly recognize that Black students, staff, and faculty have faced and continue to face unparalleled threats to their ability to succeed fully in their careers due to systemic anti-Black racism that occurs both within astronomy departments and the outside world. This reality is reflected in the data and research, as discussed above; from hostile working and learning environments to outright threats to bodily integrity, Black members of our field face disproportionate barriers to their achievement and success. Any efforts to create a more creative, productive, and equitable field must open its eyes to the blight of anti-Black racism still prevalent today. The rallying cry that "Black Lives Matter" provides AAS with an opportunity to join a growing popular movement to support Black Americans, including those who are astronomers.

Likewise, it has become unavoidably evident in 2016 that Native American nations are confronting forces of marginalization that they uniquely face, such as transgressions of their right to self-determination on lands that are still recognized as sacred. This issue of indigenous self-determination has arisen for example in the debate about the Thirty Meter Telescope, in tandem with more basic questions of what language is appropriate when describing Native Hawaiians. Moreover, Native American and Pacific Islander communities face high rates of poverty, incarceration, and violence at the hands of the police. This impacts the ability of indigenous children to be positively affected by diversity and outreach programming as well as the ability of those who are already scientists or training to become scientists to successfully pursue their professional work. How we as a professional society respond to diverse viewpoints on these concerns impacts how welcome Native American and Pacific Islander students and scientists feel in the astronomy community.

In tandem, it is especially important to recognize that transgender people, especially those who are Black, Native American, and Hispanic/Latinx, are at heightened risk for verbal abuse and violence both within the workplace and in their everyday private lives. The majority of anti-LGBTQ violence in the United States is inflicted on transgender people, with Black trans women facing some of the highest per capita murder rates in the country (NCAVP, 2016). Additionally, trans people of color are regularly criminalized for engaging in activities that are necessary for their survival. Until we can confront these realities, along with the high rates of suicide of transgender youth and adults, it will be difficult to ensure that transgender people are able to participate in astronomy. Importantly, a study released this year of LGBTQ physicists and physics students by the American Physical Society found that transgender people experienced the highest rates of exclusionary behavior in the LGBTQ community and were especially at risk for leaving the field because they had witnessed discrimination (Atherton et al., 2016).

We further recognize that in the coming years how immigration reform, bilingual education, and a host of other issues are addressed on the political stage will have a particularly strong impact on the participation of Hispanics/Latinxs in science, including astronomy. It is critical that the American Astronomical Society pro-actively engage on these issues in a way that promotes the interests and opportunities of future and current members who are affected by them. Importantly, Hispanic/Latinx people also face disproportionate rates of violence at the hands of police as well as incarceration, heightening the possibility that AAS members from this group are personally impacted by these conditions.

The creation of more diverse and inclusive environments is crucial if our field is to remain relevant, and support all members of our field in pursuing their passion and fulfilling their potential. All the data suggest that highly capable future scientists are being lost due to our failure to create warm and inclusive work environments for women, people of colour, LGBTQ people, people with disabilities, and those at the intersection of these identities. Moreover, those who stay are being subjected to non-inclusive workplace environments that cause physical and mental health problems and prevent them from reaching their full potential. As it stands, our workplaces are inadvertently hindering scientific research and standing in the way of the scientific meritocracy that we are all trying to create.

This crisis in the sciences has already been acknowledged by many different groups, including by the NSF, which is pouring $75M into new diversity initiatives; by the media, who have reported extensively on sexual harassment issues in our field; by individuals in astronomy, who have been raising these issues for decades, pushing for more inclusive policies, and organizing conferences to address inclusion in astronomy; and eventually by the AAS itself. Our field requires strong, knowledgeable, and courageous leadership to immediately implement effective solutions and programs that will support both a culture shift in individual departments and the field as a whole, and also increase the representation of underrepresented minorities among both students and faculty.

In order to know how to best support our minoritized colleagues, we need to have a better understanding of what they are experiencing on a daily basis, in both our workplaces and in society. The impacts of racism, sexism, ableism, and homophobia do not simply vanish when minoritized students, staff, and faculty enter astronomy classrooms or offices. In many cases, the interactions with our colleagues can exacerbate the harmful effects of these forms of bigotry (e.g. Cech & Waidzunas, 2011). We frequently overestimate our ability to support our minoritized students and colleagues: Samuels (2014) found that educators almost unanimously believe that they are prepared for teaching students from different cultures, but virtually none had any experience or formal training with people from other cultures. In Samuels' words, "we don't know what we don't know." If we are serious about making change and creating inclusive and diverse workplaces that support every individual, we need to get better at first knowing what we don't know and then working to fill that gap in our knowledge and understanding. We need to realize that, for people with dominant identities (i.e., men, white people, heterosexual, cissexed and cisgendered people, people who are temporarily abled), our formal education fails to educate us about the experiences of those with non-dominant identities and therefore fails to prepare us for supporting our colleagues. Our hubris is doing active damage to our minoritized students and colleagues.

One of the first steps in rectifying these discriminatory working and learning environments is to ensure that everyone in a leadership role receives diversity and inclusion training to help them understand common issues experienced by minoritized people, and to equip our leaders with the knowledge of how to further their familiarity and comprehension of these experiences (Williams, 2013). All members of the AAS Council and leadership, the heads of astronomy departments, and faculty leading in graduate selection and faculty search committees should receive both an initial, comprehensive equity and inclusion training, and also continuing education and workshops. This training should include active engagement with the latest academic research on the idea and limits of "objectivity" (Harding, 2015).

It is an unacceptable and unscientific stance to accept the status quo as the most effective way of furthering the field of astronomy. In accepting the available data and research, we put ourselves in a position to influence and encourage more equitable representation in our field. Our work on issues of diversity must be driven by best practices and research from both the academic disciplines that have studied these issues, and also from individuals and groups with demonstrated expertise in addressing these concerns. To address the pervasive inequity and bias in our community, we must recognize that most of us are amateurs in the field of diversity and equity research. Seeking out and supporting necessary training is a first step to understanding the crisis within our community. To truly make a difference, this step must be followed by active engagement with experts and those underrepresented minority members of our field to create and implement real solutions.

## Acknowledgements

Many thanks to K. Hsu for extensive suggestions and revisions, and to A. Springmann for editing and suggestions.